\documentclass{PoS}

\title{XYZ States}

\ShortTitle{}

\author{Wei Chen\\
Department of Physics, Peking University, China\\
and Department of Physics and Engineering Physics, University of
Saskatchewan, Canada}
\author{Wei-Zhen Deng\\
Department of Physics, Peking University, China}
\author{Jun He\\
Nuclear Theory Group, Institute of Modern Physics of
CAS, Lanzhou 730000, China}
\author{Ning Li\\
Department of Physics, Peking University, China\\
and Institut f\"{u}r Kernphysik and J\"ulich Center for Hadron
          Physics, Forschungszentrum J\"{u}lich}
\author{Xiang Liu\\
School of Physical Science and Technology, Lanzhou University,
Lanzhou 730000,  China}
\author{Zhi-Gang Luo\\
Department of Physics, Peking University,\\
and Department of Physics, Hubei Institute of Technology}
\author{Zhi-Feng Sun\\
School of Physical Science and Technology, Lanzhou University,
Lanzhou 730000,  China}
\author{\speaker{Shi-Lin Zhu}\thanks{Plenary talk.}\\
        Department of Physics and State Key Laboratory of Nuclear Physics and Technology \\
and Collaborative Innovation Center of Quantum Matter,\\
Peking University, Beijing 100871, China\\
        E-mail: \email{zhusl@pku.edu.cn}}

\abstract{In the past decade, many new charmonium (or
charmonium-like) and bottomonium (or bottomonium-like) states were
observed experimentally. I will review these XYZ states which do not
fit into the quark model spectrum easily.}

\FullConference{XV International Conference on Hadron Spectroscopy-Hadron 2013\\
        4-8 November 2013\\
        Nara, Japan }

\begin{document}

\section{The Charmonium Spectrum}

In the past decade many new charmonium and Upsilon (or
charmonium-like and bottomonium-like) states were observed
experimentally. Their production mechanisms include the initial
state radiation (ISR), double charmonium production, two photon
fusion, B decays and excited charmonium or bottomonium decays. These
new states were observed through either the
hidden-charm/hidden-bottom or open-charm/open-bottom final states.

Up to now the lattice QCD simulation reproduced the charmonium
spectrum below the $D{\bar D}$ threshold very well. However, many
new states above the $D{\bar D}$ threshold were observed since 2003.
Some are very narrow or even charged, which are good candidates of
exotic mesons. Beside the conventional $q\bar q$ mesons and $qqq$
baryons, QCD allows many other possible color singlets such as
dibaryons, pentaquarks, glueballs, tetraquarks, hadronic molecules,
hybrid hadrons etc.

Many new charmonium or charmonium-like states do not fit into the
quark model spectrum easily. Theoretical speculations include:
di-meson molecular states, tetraquarks, hybrid charmonium and
conventional charmonium. Molecular states are loosely bound states
composed of a pair of heavy mesons. They are probably bound by the
long-range color-singlet pion exchange. Tetraquarks are bound states
of four quarks, which are bound by colored force between quarks.
They decay through rearrangement. There are many states within the
same tetraquark multiplet. Some members are charged or carry
strangeness. Hybrid charmonium are bound states composed of a pair
of quarks and one excited gluon. Last but not the least, these new
states could still be conventional charmonium. The quark model
spectrum could be distorted by the coupled-channel effects.

Especially many charmonium-like states lie very close to the
open-charm threshold. Are these threshold enhancements real
resonances? Could they arise from the cusp effect, final state
interaction, interference between continuum and well-known
charmonium states or triangle singularity due to the special
kinematics? These possibilities always exist. One should be very
cautious in order to establish that a threshold enhancement is a
genuine resonance.

\section{The Vector Charmonium Family}

In the quark model, one expects at most five vector charmonium
states between 4 and 4.7 GeV: $3S/\psi(4040), 2D/\psi(4160),
4S/\psi(4415), 3D, 5S$. But seven states were observed
experimentally: $\psi(4008), \psi(4040), \psi(4160), \psi(4260),
\psi(4360), \psi(4415), \psi(4660)$. What are these additional Y
states?

Y(4260) was first discovered in the $J/\psi \pi\pi$ mode with the
ISR technique by Babar collaboration and confirmed by Belle and CLEO
collaboration. Y(4360) was observed in the $\psi(2S)\pi\pi$ channel
with ISR by Babar collaboration. Up to now there has been no
evidence of these two states in the open-charm process and R-value
scan. As mentioned earlier, there exist some possible non-resonant
interpretations of these states. For example, one may introduce the
interference between the continuum and the well established
resonances such as $\psi(4160)$ and $\psi(4415)$ to reproduce the
line shape of the $J/\psi \pi\pi$ and  $\psi(2S)\pi\pi$ spectrum
quite well \cite{interfere}.

It's quite possible that Y(4260) may be a conventional charmonium.
The bare $c\bar c$ state in the quark model may mix with the
$D^{(*)}{\bar D}^{(*)}$ continuum through the $D^{(*)}{\bar
D}^{(*)}$  hadron loop. The charmonium spectrum might be distorted.
For example, the screened linear potential was introduced to model
the correction from the light quark pair creation in the vacuum (or
the correction from the hadron loop). The energy level spacing above
4 GeV becomes narrower. More vector states can exist between 4 and
4.7 GeV. With such a scheme, Y(4260) can be regarded as the
$\psi(4S)$ charmonium state \cite{chao}.

Could some Y states be tetraquarks? According to the QCD sum rule
analysis, the hidden-charm vector states lie around 4.6 GeV
\cite{chenwei}. However, the hidden-charm tetraquarks will fall
apart into a pair of open-charm D mesons or hidden-charm plus light
mesons very easily. Their width is expected to be very large while Y
states are not so broad. Up to now, Y states have not been observed
in the p-wave $D^{(*)}{\bar D}^{(*)}$ modes.

If it's a genuine resonance, Y(4260) is a very good candidate of the
charmonium hybrid \cite{zhu-y4260,kou,page}. According to lattice
QCD simulation \cite{lattice}, both the vector ($1^{--}$ ) and
exotic ($1^{-+}$) hybrid charmonium lie around 4.26 GeV. One naively
expects that the vector hybrid charmonium does not couple to the
virtual photon very strongly due to the intrinsic gluon, which leads
to the dip in the R value scan. According to the flux tube model and
QCD sum rule analysis \cite{zhu-hybrid}, the favorable decay mode of
hybrid states is the p-wave + s-wave meson pair, which explains the
non-observation in the $D^{(*)}{\bar D}^{(*)}$ modes. The $c\bar c$
pair within the vector charmonium is a spin-singlet while the gluon
is color-magnetic, which is favorable to the spin-singlet
hidden-charm decay mode. Recently BESIII collaboration observed
Y(4260) in the $h_c \pi\pi$ and p-wave + s-wave D meson decay modes.

\section{The Charged States And X(3872)}

Belle collaboration reported two charged states $Z_b(10610)$ and
$Z_b(10650)$ in 2011 \cite{bellezb}. This year BESIII collaboration
observed two charged $Z_c$ states \cite{bes1,bes2}. The lower state
$Z_c(3900)$ was confirmed by Belle \cite{bellezc} and Cleoc
collaborations \cite{cleozc} very quickly.

The two $Z_c$ and $Z_b$ states are very similar. They are charged
charmonium-like (or bottomonium-like) structures close to the
open-charm (or open-bottom) threshold. Their quantum numbers are
exactly the same with $I^GJ^P=1^+1^+$. They were observed both in
the hidden-charm (or hidden-bottom) and open-charm (or open-bottom(
final states. Moreover, the open-charm (or open-bottom) modes
dominate the $Z_c$ (or $Z_b$) decays.

Could $Z_c$ (or $Z_b$) be tetraquarks? The isovector and axial
vector hidden-charm tetraquarks do lie around 4 GeV as shown in
Table V in Ref. \cite{chenwei}. However, the hidden-charm
tetraquarks will fall apart into a pair of open-charm D mesons or
one charmonium plus light mesons very easily. Their width is
expected to be large while the $Z_c$ (or $Z_b$) states are very
narrow experimentally. Moreover, the s-wave $\bar D D^*$ ($\bar B
B^*$) mode should dominate the ${\bar D}^* D^*$ (or ${\bar B}^*
B^*$) mode because of the large phase space difference. For example,
the phase space of the open-charm decay $Z_c(4020)\to {\bar D}^*
D^*$ is tiny and strongly suppressed compared to that of
$Z_c(4020)\to {\bar D} D^*$. Experimentally, the higher $Z_b(10650)$
state was not observed in the s-wave $\bar B B^*$ mode while the
higher $Z_c(4020)$ has not been observed in the s-wave $\bar D D^*$
mode. Therefore, $Z_c$ and $Z_b$ seems not good candidates of the
hidden-charm/bottom tetraquarks unless one can invent some
particular dynamics which forbids the $\bar D D^*$ and $\bar B B^*$
decay modes.

In fact, there exists a very natural interpretation. $Z_b$ states
can be regarded as candidates of the S-wave $\bar B B^*$ and ${\bar
B}^* B^*$) molecular states \cite{m1,m2,m3}. Similarly, $Z_c$ states
are candidates of the S-wave $\bar D D^*$ and ${\bar D}^* D^*$)
molecular states or molecular-type resonances \cite{m4,m5}. There
are many literatures along this direction \cite{v1,v2,hosaka}.

In QED we have the hydrogen atom where the light electron circles
around the proton. We also have the hydrogen molecule where two
electrons shared by two protons. In QCD, we have the heavy meson
where the light quark circles around heavy quark. We may also expect
the di-meson molecule where the two mesons are bound by the pion
exchange force.

The idea of the loosely bound molecular states is not new in nuclear
physics since Yukawa proposed the pion in 1935. The deuteron is a
very loosely bound state composed of a proton and neutron arising
from the color-singlet meson exchange. Besides the long-range pion
exchange, the medium-range attraction from the correlated two-pion
exchange (or in the form of the sigma meson exchange), the
short-range interaction in terms of the vector meson exchange, and
the S-D wave mixing combine to form the loosely bound deuteron. The
deuteron is a perfect hadronic molecular state!

We adopt the same one-boson-exchange formalism to discuss the
possible molecular states composed of a pair of heavy mesons. Within
this framework, both $Z_b(10610)$ and $Z_b(10650)$ can be explained
as the S-wave $\bar B B^*$ and ${\bar B}^* B^*$) molecular states.
Besides the isovector $Z_b$ states, there are also several loosely
bound isoscalar molecular states \cite{m1}. The same analysis holds
for the $Z_c$ system \cite{m4,m5}.

Then how about X(3872)? In Refs. \cite{lining-x3872,m2,m3}, we
considered (i) the S-D wave mixing which plays an important role in
forming the loosely bound deuteron; (ii) both the neutral ${\bar
D}^0 D^{\ast 0}$  and charged ${\bar D}^+ D^{*-}$ component in the
flavor wave function; (iii) the mass difference between the neutral
and charged $\bar D D^*$ meson; and (iv) the coupling of $\bar D
D^*$ to ${\bar D}^* D^*$ channel. It turns out that X(3872) is a
very loosely bound molecular state. When the binding energy is 0.3
MeV, the branching fraction ratio between the isospin conserving
$J/\psi \pi^+\pi^-\pi^0$ mode and isospin violating $J/\psi
\pi^+\pi^-$ is 0.42 \cite{lining-x3872}, which agrees well with both
Babar's measurement $0.8\pm 0.3$ and Belle's measurement $1.0\pm
0.4\pm 0.3$. It's important to note that we do not need to add by
hand the $J/\psi \rho$ and $J/\psi \omega$ component into the flavor
wave function of X(3872) in order to explain the large isospin
violation in its strong decays.

The existence of the loosely bound state X(3872) and the large
isospin symmetry breaking in its hidden-charm decay arises from the
very delicate efforts of the several driving forces including: the
long-range one-pion exchange, S-D wave mixing, mass splitting
between the charged and neutral $D(D^*)$ mesons and coupled-channel
effects. The extreme sensitivity of the physical observables to the
tiny binding energy is typical of the loosely bound system.

Then how can we further test the one-pion-exchange model and
molecular picture? We can test the molecular picture through the
isoscalar partner of $Z_c(3900)$, isovector partner states in the
$Z_c(4020)$ multiplet with $I^G(J^P)=1^- (0^+), 1^+(1^+), 1^-
(2^+)$, and isoscalar partner states of $Z_c(4200)$ with $I^G
J^{PC}=0^+(0^{++}), 0^-(1^{+-}), 0^+(2^{++})$. One may wonder
whether X(3872) is the isoscalar partner of $Z_c(3900)$. We can also
investigate the pionic, di-pion and electromagnetic decay pattern of
the $Z_(3900)$ and $Z_c(4020)$ multiplet with the heavy quark spin
and flavor symmetry. Details can be found in Ref. \cite{m4}. Similar
discussions hold for the $Z_b$ states.

\section{The Possible Landscape of Hadronic Molecules}

This year BESIII collaboration reported the hidden-charm decays
$Y(4260)\to \pi Z_c(3900, 4020)$, $Y(4360)\to \pi Z_c(3900, 4020)$
and radiative decay $Y(4260)\to \gamma X(3872)$ with a large ratio
$\frac{\Gamma [\gamma X(3872)]}{ \Gamma [ \pi\pi J/\psi]} =11.2\%$
\cite{bes3}. It seems that there might exist some possible
connections between various XYZ states. Are they related to each
other? Could they have the same inner structure?

Theoretical speculations of Y(4260) include (a) $\psi (4S)$; (b) a
vector hybrid charmonium; (c) a $D_1\bar D$ molecule. In contrast,
X(3872) could be (1) $\chi_{c1}^\prime$; (2) a $\bar D D^*$
molecule; (3) the mixture of $\chi_{c1}^\prime$ and $\bar D D^*$
molecule. There are many combinations. In the following let's
discuss several typical scenarios.
\begin{itemize}
\item*[Scenario 1:]\\
Y(4260)=$\psi (4S)$ and X(3872)=$\chi_{c1}^\prime$. Then we expect
no more charged partner states around Y(4260)? It's very challenging
to explain why Y(4260) was not observed in p-wave ${\bar D}(*) D(*)$
modes? It's also very difficult to explain the large $J/\psi\pi\pi$
branching ratio of Y(4260)? For comparison, the $J/\psi\pi\pi$
branching ratio of well established excited charmonium is around
$1\%$. Moreover, where is the isoscalar partner of $Z_c(3900)$ if
it's not an experimental fake signal?

\item*[Scenario 2:] \\
Y(4260)=$\psi (4S)$ and X(3872)=$\bar D D^*$ molecule. Again, the
same puzzle of Y(4260) exists as in Scenario 1. Then where is
$\chi_{c1}^\prime$?

\item*[Scenario 3:] \\
Y(4260)=hybrid charmonium and X(3872)=$\chi_{c1}^\prime$. There are
no more charged partner states around Y(4260). Then what is the
similar state Y(4360)? Where is the isoscalar partner of
$Z_c(3900)$?

\item*[Scenario 4:] \\
Y(4260)=$D_1\bar D$ molecule and X(3872)=$\bar D D^*$ molecule. This
scenario is both interesting and a little wild. There will be many
charged partner states around Y(4260). Y(4360) may also be a
molecular state. With Scenario 4, we expect a landscape of hadronic
molecules.
\end{itemize}

First we consider molecules composed of two S-wave D mesons. Similar
discussions can be easily extended to the B meson system.
\begin{itemize}
\item ${\bar D} D^*[I=0, J=1]$ =X(3872)
\item ${\bar D} D^*[I=1, J=1]$ = $Z_c(3900)$ multiplet
\item ${\bar D}^* D^*[I=0, J=0,1,2]$ = missing
\item ${\bar D}^* D^*[I=1, J=1]$=$Z_c(4020)$ multiplet \\
Maybe Belle's $Z^+(4051)$ observed in the $\pi\chi_{c1}^\prime$
channel is a member of this multiplet?
\item ${\bar D}^* D^*[I=1, J=0]$=missing?
\item ${\bar D}^* D^*[I=1, J=2]$=missing or does not exist?
\item Strange or hidden-strangeness partners?
\end{itemize}

There are also possible molecules composed of one S-wave D meson and
another P-wave D meson.
\begin{itemize}
\item $D_1 {\bar D}[I=0, J=1]$ =Y(4260)
\item $D_1 {\bar D}[I=1, J=1]$ =?\\
Maybe Belle's $Z^+(4248)$ observed in the $\pi\chi_{c1}^\prime$
channel is a member of this multiplet?
\item $D_1 {\bar D}^*[I=0, J=1]$ =Y(4360)?
\item $D_1 {\bar D}^*[I=0, J=0,2]$ = missing
\item $D_1 {\bar D}^*[I=1, J=0,1,2]$ =?\\
Maybe Belle's $Z^+(4430)$ observed in the $J/\psi\pi$ channel is a
member of this multiplet \cite{z4430a,z4430b}?
\item $D_2 {\bar D}[I=0,1; J=2]$ =missing or does not exist?
\item $D_2 {\bar D}^*[I=0,1; J=1,2,3]$==missing or does not exist?
\item Strange or hidden-strangeness partners?\\
Maybe Y(4660) is an isoscalar hidden-strangeness member belonging to
this group?
\end{itemize}

\section{Summary}

The excited charmonium or Upsilon states act as a hadron molecule
factory. The $Y(5S)$ mass is 10.86 GeV while
$M_B+M_{B^*}+M_\pi=10.744$ GeV and $2M_{B^*}+M_\pi=10.79$ GeV. The
phase space of the $Y(5S)\to B(*){\bar B}(*)\pi$ decay is tiny. The
relative motion between the B meson pair is very slow, which is
favorable to the formation of the molecular states. $Y(5S, 6S)$ is
the ideal factory of molecular states. These molecular states will
be produced abundantly at BelleII in the near future!

Similarly, the excited charmonium decay such as Y(4260, 4360, 4660)
etc is ideal in the search of the D meson molecular states. The
$\gamma$, $1\pi$, $2\pi$, $3\pi$ and other light degree of freedom
will act as a quantum number filter of these interesting states.

X(3872), $\chi_{c1}^\prime$ and Y(4260) are the key states in
revealing the underlying structure of the charmonium-like XYZ
states. Through the decay pattern and possible partner states, we
can test the molecule picture. Especially the experimental
measurement of the various pionic and electromagnetic transitions
between Y(4360), Y(4260), $Z_c(4020)$, $Z_c(3900)$ and X(3872) are
crucial.

\section*{Acknowledgments}

This project was supported by the National Natural Science
Foundation of China under Grants 11075004, 11021092, and
11261130311.

\end{document}